\documentclass{aa}
\usepackage{graphicx}
\usepackage{times}
\usepackage{mathptm}
\fontfamily{ptmcm}\selectfont
\def\ltsim{\raise 2pt \hbox {$<$} \kern-1.1em \lower 4pt \hbox {$\sim$}}
\def\gtsim{\raise 2pt \hbox {$>$} \kern-1.1em \lower 4pt \hbox {$\sim$}}
\input epsf.sty
\begin{document}
%   \thesaurus{13(13.18.1); % Radio continuum: galaxies
%              11(11.09.1;       % {\bf Galaxies: individual:}   
%           11.19.6;      %Galaxies: structure
%               11.17.4; % {\it (Galaxies:)} {\bf quasars: individual:}
%                 11.14.1)} %Galaxies: nuclei
%       

   \title{Radio monitoring of a sample of X-- and $\gamma$--ray loud blazars}

\author{T. Venturi
\inst{1}, D. Dallacasa\inst{2}, A. Orfei\inst{1}, M. Bondi\inst{1},
R. Fanti\inst{1,3}, L. Gregorini\inst{1,3}, F. Mantovani\inst{1},
C. Stanghellini\inst{4}, C. Trigilio\inst{4}, G. Umana\inst{4}}

\offprints{T. Venturi(tventuri@ira.bo.cnr.it)}

\institute{1. Istituto di Radioastronomia del CNR, Via Gobetti 101, 
40129 Bologna, Italy\\
2. Dipartimento di Astronomia, Bologna University, Via Ranzani 1, 
40127 Bologna, Italy\\
3. Dipartimento di Fisica, Bologna University, Via Irnerio, 46, 40126
Bologna, Italy\\
4. Istituto di Radioastronomia del CNR, C.P. 141, Noto (SR), Italy}

\titlerunning{Radio monitoring of blazars}
\authorrunning{Venturi T. et al.}

\date{Received ...; accepted ...}

%\markboth{Venturi T. et al.: Radio monitoring of blazars}
%{T.Venturi et al.: Radio monitoring of blazars}

\abstract
{
In this paper we present the results of a 4-year (1996 - 1999)
radio flux density monitoring program for a sample of X-- and
$\gamma$--ray loud blazars.
Our program started in January 1996 and was carried out on monthly 
basis at the frequencies of 5 GHz and 8.4 GHz with the 32-m 
antennas located in Medicina (Bologna, Italy) and Noto (Siracusa, Italy).
%5 GHz observations before October 1997 were carried out
%with the 32-m antenna located in Noto (Siracusa, Italy).
22 GHz data collected in Medicina from January 1996 to June 1997
will also be presented. 
The sample of selected sources comprises most radio loud blazars
with $\delta \ge -10^{\circ}$ characterised by emission in the
X-- and $\gamma$--ray regimes, and target sources for the BeppoSAX
X--ray mission. All sources in the sample, except J1653+397 (MKN\,501),
are variable during the four years of our monitoring 
program. We classified the type of variability in each source by means 
of a structure function analysis. We also computed the
spectral index $\alpha_{5}^{8.4}$ for all epochs with nearly 
simultaneous observations at these two frequencies, and found that 
$\alpha_{5}^{8.4}$ starts flattening at the very beginning of a radio flare,
or flux density increase.
}
%
%  14.Sep.'90: Demo-Vs.
%________________________________________________________________
   \maketitle

\section{Introduction}

The flux density variability of compact extragalactic radio 
sources is a well established phenomenon, which can be explained 
as due to the propagation of shocks in relativistic jets 
aligned close to the observer's line of sight (see for example Marscher
\& Gear 1985). The large amount
of data available in a wide range of radio frequencies, makes
it clear that different behaviours for the radio variability in blazars
exist, and many different classes of spectral variability can
be identified (for a review see Marscher 1993 and Wagner \& Witzell 1995,
and references therein).

\noindent
The variability of blazars at X-- and $\gamma$--ray energies is not
as well documented as the radio flux density variability, however 
a study of simultaneous X--ray and radio--infrared flares
has been reported for a few objects. Some examples
include 3C279 (Maraschi et al. 1994, Wehrle et al. 1998), 
BL Lac (Kawai et al. 1991), NRAO 140 (Marscher 1988), PKS 2255$-$282
(Tornikoski et al. 1999), MRK 501 (Petry et al. 2000).

There are at present
two main models accounting for the correlation observed.
In the accelerating jet model (Maraschi et al. 1992)
the synchrotron emission at UV, optical and IR frequencies is confined
to the region closest to the central engine, opaque to the radio emission.
The radio emission is produced outside this region, with the maximum
intensity occurring where the Lorentz factor $\Gamma$ is highest.
Following this model,
self-Compton scattered $\gamma$-- and X--rays are produced 
in coincidence with the synchrotron emission in the UV, optical
and IR bands closest to the central engine and in the radio core.
Moreover, Inverse Compton reflection of 
optical and UV photons produced by the accretion disk would take place
in the vicinity of the central engine, again producing X-- and
$\gamma$--ray emission.
In a model consisting of a decelerating flow of relativistic positrons
and electrons (Melia and Konigl 1989) the UV photons
produced by the accretion disk are upscattered to X- and $\gamma~$- ray
energies. Radio and infrared synchrotron emission (plus self-Compton
scattered X-- and $\gamma~$--ray emission) is produced where the Lorentz factor
decreases down to a value of $\sim$ 10.
Both models connect the emission in the different energy bands, and 
the location along the jet where the emission takes place, so
in principle it should be possible to discriminate among models and
throw a light on the nature of the inner jets in blazars by means
of multifrequency observations (Marscher 1993).

In order to increase the number of blazars (i.e. BL Lacertae objects, 
optically violent variables, high and low polarisation quasars)
with multifrequency studies, we selected a sample of 23 radio
loud X-- and $\gamma$--ray blazars, targets of X--ray observations
carried out with the satellite BeppoSAX, and observed them 
at centimeter wavelengths on monthly basis, for comparison with the
results of the BeppoSAX X--ray mission.
Our aim was to observe all sources in our list to provide radio
lightcurves with very good and regular time coverage, which is 
crucial for an accurate multifrequency study of their variability. 
The desire for simultaneous (or nearly simultaneous) observations
in the various bands is a result of the very high variability of these 
sources over the whole electromagnetic spectrum, and of the physical 
process responsible for the radio emission.

We report the source list, the
description of the adopted observing strategy and the results 
in Section 2. A brief analysis of the variability
and of the spectral index behaviour for all sources in the
sample will be given in Section 3.

\section{Source selection and observations}

\subsection{The sample of radio loud $\gamma$--ray blazars}

The objects in our sample were selected from the list
of steep spectrum and flat spectrum X--ray blazars,
targets of observations with the X--ray satellite BeppoSAX.
We chose all blazars with $\delta \ge -10^{\circ}$ 
with radio flux density S$~\ge$ 1 Jy at 5 GHz at the time we started our
program. We point out that because of the strong 
variability of this class of sources, the list of selected
objects cannot be considered complete.

Our sample includes 23 radio sources, reported 
in Table 1 with the most relevant informations, i.e. J2000 and 
B1950 IAU  name (column 1);
alternative name (column 2); accurate J2000 radio coordinates taken 
from the VLA
calibrator list (columns 3 and 4); redshift (column 5); optical ID (column 6).
All sources in our list are part of the UMRAO 
database,
see for example Aller et al. (1985) and the web page address:

\centerline{http://www.astro.lsa.umich.edu/obs/radiotel/umrao.html}

\noindent
and have 
been monitored since the early seventies at centimeter wavelenghts.
However, our monitoring program presented 
in the next section, does not duplicate but complements the
major observational effort of the University of Michigan Radio Astronomy
Observatory, especially in the light of the multiband analysis
which motivated our observational project. 

\noindent
Five sources in our list were monitored with the Green Bank 
Interferometer
at 8.4 GHz and 2.7 GHz until October 2000. Lightcurves and flux densities 
are available on web at the address: 
http://www.gb.nrao.edu/fgdocs/gbi/gbint.html.
%
% begin table 1
\begin{table*}

\caption{The radio sources}

\begin{tabular}{llcrcc} \hline
$~~~~~~~~~~$IAU Name & Other Name & RA (J2000) &  DEC (J2000) &  z  &  
Opt. ID \\
  &   &  &  &  &   \\ \hline
J0050$-$094$~~$ 0048$-$097&  & 00 50 41.32 &$-$09 29 05.2 &  -    & BL-Lac \\
J0238+166  $~~$ 0235+164  &  & 02 38 38.93 &   16 36 59.3 & 0.940 & BL-Lac \\
J0530+135  $~~$ 0528+134  &  & 05 30 56.42 &   13 31 55.1 & 2.060 &  QSO   \\
J0721+713  $~~$ 0716+714  &  & 07 21 53.45 &   71 20 36.4 &  -    &  QSO   \\
J0738+177  $~~$ 0735+178  &  & 07 38 07.39 &   17 42 18.9 & 0.424 & BL-Lac \\
J0841+708  $~~$ 0836+710  &  & 08 41 24.37 &   70 53 42.2 & 2.172 &  QSO   \\
J0854+201  $~~$ 0851+202  &OJ\,287& 08 54 48.87 &20 06 30.6 & 0.306 & BL-Lac \\
J0958+655  $~~$ 0954+658  &  & 09 58 47.25 &   65 33 54.8 & 0.368 & BL-Lac \\
J1104+382  $~~$ 1101+384  &MKN\,421&11 04 27.31 &38 12 31.8 & 0.030 & BL-Lac \\
J1229+020  $~~$ 1226+023  &3C\,273& 12 29 06.70 &02 03 08.6 & 0.158 &  QSO   \\
J1256-057  $~~$ 1253$-$055&3C\,279&12 56 11.17 &$-$05 47 21.5 & 0.536 &QSO   \\
J1419+543  $~~$ 1418+546  &OQ\,530& 14 19 46.60 &54 23 14.8 & 0.151 & BL-Lac \\
J1512-090  $~~$ 1510$-$089&  & 15 12 50.53 &$-$09 05 59.8 & 0.360 &  QSO   \\
J1653+397  $~~$ 1652+398&MKN\,501& 16 53 52.22 & 39 45 36.6 & 0.034 & BL-Lac \\
J1748+700  $~~$ 1749+701  &  & 17 48 32.84 &   70 05 50.8 & 0.770 & BL-Lac \\
J1751+096  $~~$ 1749+096  &  & 17 51 32.82 &   09 39 00.7 & 0.322 & BL-Lac \\
J1800+784  $~~$ 1803+784  &  & 18 00 45.68 &   78 28 04.0 & 0.680 & BL-Lac \\
J1806+698  $~~$ 1807+698&3C\,371& 18 06 50.68 &69 49 28.1 & 0.051 & BL-Lac \\
J1824+568  $~~$ 1823+568&4C\,56.27&18 24 07.07&56 51 01.5 & 0.664 & BL-Lac \\
J2005+778  $~~$ 2007+777  &  & 20 05 30.99 &   77 52 43.2 & 0.342 & BL-Lac \\
J2202+422  $~~$ 2200+420&BL\,Lac& 22 02 43.29 &42 16 39.9 & 0.069 & BL-Lac \\
J2232+117  $~~$ 2230+114&CTA\,102& 22 32 36.41 &11 43 50.9 & 1.037 &  QSO   \\
J2253+161  $~~$ 2251+158&3C\,454.3&22 53 57.75 &16 08 53.6 & 0.859 & QSO   \\
\hline

\end{tabular}
\end{table*}

\subsection{Observational strategy, data reduction and results}

We started our monthly monitoring program at 8.4 GHz (3.6 cm) and 5 GHz
(6 cm) in January 1996. All the 8.4 GHz observations were carried out
with the 32-m antenna located in Medicina (Bologna, Italy);
the 5 GHz observations prior to October 1997 were carried out 
with the 32-m  antenna located in Noto (Siracusa, Italy), while 
the later 5 GHz observations were carried out in Medicina.
The observations at 5 GHz and 8.4 GHz after this date were typically
carried out within a few days one another.
The gap in the 8.4 GHz data, from June to November 1996, is due to
the antenna track repair in Medicina.
We present here also 22 GHz (1.3 cm) observations carried out in 1996 and
part of 1997 in Medicina for the same set of sources.

Both left and right circular polarisation were recorded in Medicina, 
with a bandwidth of 80 MHz around the central frequencies
4967 MHz, 8447 MHz and 22331MHz. 
Total band observations, i.e. 4700 - 5050 MHz, 
were carried out in Noto, where only the left circular
polarisation was recorded. The features of
the antenna and of the cooled receivers at these frequencies 
are reported in 
Table 2, where we give the half power beamwidth (HPBW), the
peak gain ({\it g}) and the receiver temperature system at
the zenith (T$_{sys}$) at each observing frequency. We note that the
zenith T$_{sys}$ includes also the sky.

In order to account for the local background,
the data acquisition was performed with the on-source/off-source 
method, by means of the program 
ON$-$OFF. The off-source measurements were done 5 beams off-source 
(on both sides of the source), in the azimuth direction. 
The duration of each ON$-$OFF cycle varied from 15 minutes for the 
strongest  sources to 75 minutes for the weakest, which implies an 
effective integration time on source ranging from $\sim 3$ to 
$\sim 15$ minutes if we take into account all phases of the cycle.

DR21 was used as primary calibrator at both frequencies, with 
adopted flux densities of  S$_{5 GHz}$ = 22.5 Jy (Baars et al. 1977) and 
S$_{8.4 GHz}$ = 21.5 Jy (Ott et al. 1994),
S$_{22 GHz}$ = 17.0 Jy (Baars et al. 1977).
We observed it 6$-$7 times each
observing run, in the range of elevations $20^{\circ} - 85^{\circ}$.
3C123, 3C286 and 3C274 were used as secondary calibrators and
were observed twice each run.

% begin table 2
\begin{table*}

\caption{Parameters of the observations}

\begin{tabular}{lrcccc} \hline
Antenna  & $\nu$& Obs. Period & HPBW  & {\it g} (L,R) & T$_{sys}$ (L,R) \\
         &  MHz &   mm/yy     & arcmin&       K/Jy    &   K        \\ \hline
Medicina & 4967 &11/97 - 01/00&  7.5 & 0.160$~~$0.161 & 46$~~$52   \\
         & 8447 &01/96 - 01/00&  4.8 & 0.145$~~$0.135 & 39$~~$37   \\
         & 22331&01/96 - 06/97&  2.0 & 0.116$~~$0.118 & 120$~~$120 \\  
Noto     & 4875 &01/96 - 10/97&  7.5 & 0.161$~~~~~~~~$& 45$~~~~~$  \\ \hline

\end{tabular}
\end{table*}
\noindent

The data reduction was carried out for the two polarisations 
independently using the program CINDY (Sanfilippo,
Trigilio \& Umana 1995), written specifically for ``single dish''
observations. The antenna temperature on source T$_{ant}$, 
derived on the basis of the on-source and off-source measurements, 
is transformed into flux density S by means of the formula: 
S = T$_{ant}$ / G(z), where G(z) is the antenna gain as function of
the source zenith angle. The data points were averaged to obtain
a single flux density value, and the flux densities from the
two polarisations were subsequently averaged together. We assumed
no flux density variations during each observation.

\medskip
The calibration uncertainty in our data is of the order of
4\% at all frequencies. This value was estimated comparing all the 
gain curves determined for each observing run.

\noindent
The expected thermal noise {\it rms} in our observations (see above and
Table 2) is of the order of 2 - 5 mJy at 5 GHz and 8.4 GHz,
and 6 - 15 mJy at 22 GHz.
The most serious source of error in this type of observations
is the uncertainty in the estimate of the atmospherical optical
depth. The on-off procedure allows the subtraction of the ``local''
sky, therefore each individual point in the observations has already
been corrected for this effect. We note, however, that the
non-simultaneous measurements on- and off-source introduce an
additional uncertainty, which we empirically estimated to increse the
expected thermal noise by a factor of $\sim$ 4.

The results of our monitoring program are reported in Table 3, 
where we give the epoch and the flux
density in Jy at 8.4 GHz, 5 GHz and 22 GHz. The error associated
to each datapoint can be estimated following the indications given
in the previous paragraphs with the simple formula:

\centerline{$\Delta$S = 0.04$\times$S + 4$\times$ {\it rms}.}

\noindent
In Figure 1 we show the lightcurves at all frequencies,
and the spectral index between 8.4 GHz and 5 GHz for those 
epochs with time separation $\ltsim~$ 7 days. 
The error bars in the plots take into account all the
sources of error illustrated above.

Our data are in very good agreement with the flux density
measurements provided by the UMRAO database for the common epochs. 
For the five sources included in the GBI sample, we found good 
consistency between ours and the GBI data.

\section{Data Analysis}

\subsection{Comments on the flux density variability}

As it is clear from Figure 1 and Table 3, all sources in the sample show some 
degree of flux density variability at centimeter wavelengths during
the four years of our monthly monitoring.
The most quiescent source is J1653+397 (MKN\,501), whose light curve
remains almost flat both at 5 GHz and 8.4 GHz. From our data it is clear
that the major outburst at X--ray energies detected for this source in 1997 
did not propagate down to the radio regime covered by our observations
(see also Petry et al. 1998). As discussed by Pian et al. (1997),
the spectral properties of the 1997 flare and the major shift of
the synchrotron peak, challenge most jet models.

We carried out a simple classification of the variability in all sources,
applying a structure function analysis (Hughes, Aller \& Aller 1992) to 
the 8.4 GHz observations. This dataset is the most complete 
among those presented in this paper, and for this reason it is
particularly suited for this study.
 
The structure function is defined as $SF(\tau) = <[S(t) - S(t+\tau)]^2>$, 
where  $S(t)$ is the flux at the time $t$ and $\tau$ is the time lag.
For an ideal process $SF$ increases until it reaches a plateau,
starting at a time $T_{max}$, which corresponds to the timescale of the 
variability. The real cases, however, usually give more complicated 
structures. 

We identified three different classes of variability, briefly described in 
the following paragraph. The variability class 
for each source is given in column 1 of Table 3. In Fig. 1, lower 
panel, we report the SF for each source.

\noindent
{\it Class (a) -} Structure functions without a clearly defined plateau.
This suggests that the time scale of the variability is longer than the 
duration of our monitoring.
Examples in this class are J0530+135 and J1824+568 (4C\,56.27). 
As visible from the light curve 
plots given in Figure 1, some modulations are actually present in all 
these sources, however it is possible that their low amplitude 
compared to the global trend, was missed by the analysis.

\noindent
{\it Class (b) -} Structure functions with both axima and minima.
This case corresponds to
recurrent variability, and the clearest examples are J0050$-$094, J1419+543 
(OQ\,530) and J1800+784.
In all sources belonging to this class the amplitude of the variability 
is similar during the four years of monitoring.

\noindent
{\it Class (c) -} Structure functions with one or more  
plateau.
The first case, which includes for example J0238+166, identifies
sources with one single outburst, while the presence of more
than one plateau reflects the existence of different types of 
variability in the same source, such as for example in 
J0958+655, J1751+096 and J2253+161 (3C\,454.3). In these sources short term 
flux density fluctuations and longer term variations seem to overlap.
The majority of the sources in our sample belongs to this class.

We underline that there seems to be no difference in the structure 
function of quasars and BL Lacs, in that both populations include 
objects in each variability class.

\medskip
Comparison of the flux density variability at 22 GHz, 8.4 GHz and 5 GHz 
in each source (see Figure 1), 
indicates that not all sources in the sample show the
same behaviour. In most cases the variability
in these bands overlap, with simultaneous, or nearly
simultaneous maxima and minima. This is particularly clear for the
recurrent variability of J0050$-$094, J1419+543 (OQ\,530)
and of J1800+784, but 
also for  non periodic variable sources such as J0530+135 and J1104+382
(MKN\,421). In other cases, as for example J0238+166, J1229+020 (3C\,273),
J1256$-$057 (3C\,279) and J1751+096, the 
flux density increase is visible first at higher frequencies, i.e.
22 GHz and 8.4 GHz, and subsequently
propagates to 5 GHz. The time delay between the two
frequencies seems to be of the order of a month for J0238+166
and J1751+096, and of the order of few months for the other two sources.

One possible explanation to account for this result is that
the mechanism responsible for the variability differs in the various 
cases. An analysis of the correlation between 
flux density variations and nuclear (i.e. parsec-scale) morphological 
changes, was carried out by Zhou et al. (2000) for the blazar
PKS\,0420$-$014. For that source it was proposed that radio
outbursts with time delay going from high to low radio frequencies 
are due to intrinsic variations in the source nucleus, while
simultaneous flares at different frequencies may be due to
geometric effects. It would be important to carry out such 
analysis for a larger sample of sources.

\medskip
On the basis of the lightcurves derived from our monitoring program,
a subsample of sources was selected for parsec-scale imaging and further
investigation of their nuclear properties.
In particular, J0050$-$094, J0238+166, J0958+655, J1512$-$090, J1751+096
were selected as representatives of different classes of variability
(see the classification in Table 3),
and multiepoch observations were carried out 
with the Very Long Baseline Array (VLBA) at 8.4 GHz and 22 GHz,
yielding an angular resolution $\ltsim$ 1 mas.
The observations were performed on 22 Jan 1999 and on 7 Dec 2000.
In addition, J1512$-$090 was observed with the VLBA and 
the Space VLBI antenna HALCA on 11 Aug 1999 and 13 May 2000, with
a resolution comparable to that of the 22 GHZ ground-array
observations.

\noindent
J0050$-$094 and J0238+166 were found to be unresolved at both
frequencies, while superluminal motion was found in J0958+655 and J1512$-$090,
with implied Lorentz factors in the range 1.7 - 5.
Preliminary results are presented in Venturi et al. 2000 and 2001, and 
further analysis is in progress (Venturi et al., in preparation).

\subsection{Comments on the spectral index}

The time coverage at 8.4 GHz and 5 GHz allowed
the computation of the spectral index $\alpha_5^{8.4}$ 
for a number of epochs, especially starting from 
November 1997, when observations at both frequencies were carried 
out nearly  simultaneously with the 32-m Medicina radio telescope.
The trend of $\alpha_5^{8.4}$ versus epoch, to be read in the sense 
S$\propto \nu^{-\alpha}$, is reported in the bottom frame of the
upper panel of Fig. 1.

As typical for compact radio sources, the spectral index is flat,
i.e. $\alpha_5^{8.4} \sim 0$, or inverted in all cases, and it is
mostly in the range $-0.5~$ \ltsim $~\alpha_5^{8.4}$ \ltsim 0.
The only two cases where $\alpha_5^{8.4}~$ \gtsim 0 during the
whole duration of our monitoring are J1653+397 (MKN\,501) and 
J2253+161 (4C\,454.3).

The spectral index usually starts steepening at the very beginning of 
a radio flare, or flux density increase, as clear in particular for 
J0238+166 and J1229+020 (3C\,273). This reflects the fact that 
the flux density increase takes place first at high frequencies, and/or
that the amplitude of the flux density variations is on average more
pronounced at 8.4 GHz.

\section{Summary}

We presented the results of a 4-year radio monitoring program carried 
out for 23 radio loud blazars. In particular, the selected sources in
the sample were monitored on monthly basis at 8.4 GHz and 5 GHz, 
from January 1996 to January 2000. Observations at 22 GHz in the period
January 1996 - June 1997 are also shown.
The original aim was to provide good radio lightcurves for a set of 
blazars observed with the X--ray satellite SAX, for multiband analysis
and to determine the spectral energy distribution (SED) from
radio to X-- and $\gamma$--ray energies for a larger number of
such objects.

With the exception of J1653+397 (MKN\,501), all sources in our sample
showed some degree of variability during the period of the monitoring.
A structure function analysis suggests that not all sources in
the sample are characterised by the same type of variability. 
In some cases proper
flux density flares are detected; in others recurrent variability
is seen, and finally a more complicated behaviour is revealed in a number 
of sources, with flares of small amplitude superposed on longer
tem variations. 
A detailed study of the lightcurves presented here will be done 
together with the X--ray data from BeppoSAX in a forthcoming paper.
In particular, a thorogh study and multiband analysis of the five sources 
imaged at parsec-scale resolution is in progress.

%%%%%%%%%%%%%%%

\begin{acknowledgements}

This work has made use of the NASA/IPAC Extragalactic Database (NED),
and of data from the University of Michigan Radio Astronomy,
which is supported by the University of Michigan.

\end{acknowledgements}

\clearpage

% begin table 3
\begin{table*}

\caption{Flux density measurements (Sample page. The complete table can
be found at the CDS)}

\begin{tabular}{lcccccc} \hline
Source & Date       &  S (Jy) & Date & S (Jy) & Date
& S (Jy) \\
Var. Class& y\,m\,d & 8.4 GHz & y\,m\,d  & 5 GHz & y\,m\,d  & 22 GHz \\ \hline
{\bf J0050$-$094} & 1996\,01\,30   &  1.42  &              &          & 
1996\,02\,06   &  1.63     \\
{\it (b)} &1996\,03\,04   &  1.63     &               &            &
1996\,03\,09   &  1.70     \\
& 1996\,03\,29   &  1.92     &  1996\,04\,15 & 1.29   &
1996\,03\,30   &  1.63     \\
& 1996\,05\,06   &  1.57     &               &            & & \\
& 1996\,05\,24   &  1.52     &               &            &
1996\,05\,25   &  1.57     \\
& 1996\,11\,20   &  1.76     &  1996\,11\,13 & 1.60   & & \\  
& 1997\,01\,29   &  1.94     &  1997\,02\,06 & 1.43   &
1997\,01\,14   &  1.88     \\  
& 1997\,03\,25   &  1.50     &               &            &
1997\,03\,22   &  1.19     \\  
& 1997\,04\,15   &  1.38     &               &            & & \\
& 1997\,05\,12   &  1.26     &               &            & & \\
& 1997\,06\,20   &  1.41     &               &            & & \\
& 1997\,07\,26   &  1.68     &               &            & & \\
& 1997\,08\,01   &  1.70     &               &            & & \\
& 1997\,08\,17   &  1.81     &  1997\,10\,11 & 1.60   & & \\  
& 1997\,08\,29   &  1.83     &  1997\,12\,24 & 1.54   & & \\  
& 1997\,10\,04   &  2.00     &  1998\,01\,14 & 1.22   & & \\  
& 1998\,02\,07   &  1.17     &  1998\,02\,04 & 1.13   & & \\  
& 1998\,03\,04   &  1.21     &  1998\,03\,07 & 1.00   & & \\  
& 1998\,03\,28   &  1.47     &               &            & & \\
& 1998\,05\,07   &  1.55     &  1998\,05\,05 & 1.40   & & \\  
& 1998\,06\,27   &  1.45     &               &            & & \\
& 1998\,07\,17   &  1.23     &               &            & & \\
& 1998\,07\,29   &  1.17     &               &            & & \\
& 1998\,09\,04   &  1.17     &  1998\,08\,31 & 1.06   & & \\
& 1998\,10\,09   &  1.49     &  1998\,10\,02 & 1.33   & & \\  
& 1998\,12\,05   &  1.85     &  1998\,12\,07 & 1.62   & & \\  
& 1999\,01\,16   &  2.06     &  1999\,01\,05 & 1.91   & & \\  
& 1999\,02\,06   &  2.05     &               &            & & \\
& 1999\,05\,05   &  1.42     &               &            & & \\
& 1999\,06\,25   &  1.06     &               &            & & \\
& 1999\,07\,15   &  1.03     &  1999\,07\,17 & 0.97   & & \\  
& 1999\,07\,22   &  1.02     &  1999\,07\,20 & 0.97   & & \\  
& 1999\,08\,18   &  1.22     &  1999\,08\,20 & 1.04   & & \\  
& 1999\,10\,07   &  1.16     &  1999\,10\,01 & 1.12   & & \\  
& 1999\,11\,06   &  1.19     &  1999\,11\,04 & 1.15   & & \\  
& 1999\,12\,06   &  1.36     &               &            & & \\
& 2000\,01\,28   &  1.55     &  2000\,01\,10 & 0.95   & & \\  \hline
\end{tabular}
\end{table*}

\clearpage
% figure 1 
\setcounter{figure}{0}
\begin{figure*}
\resizebox{\hsize}{!}{\includegraphics{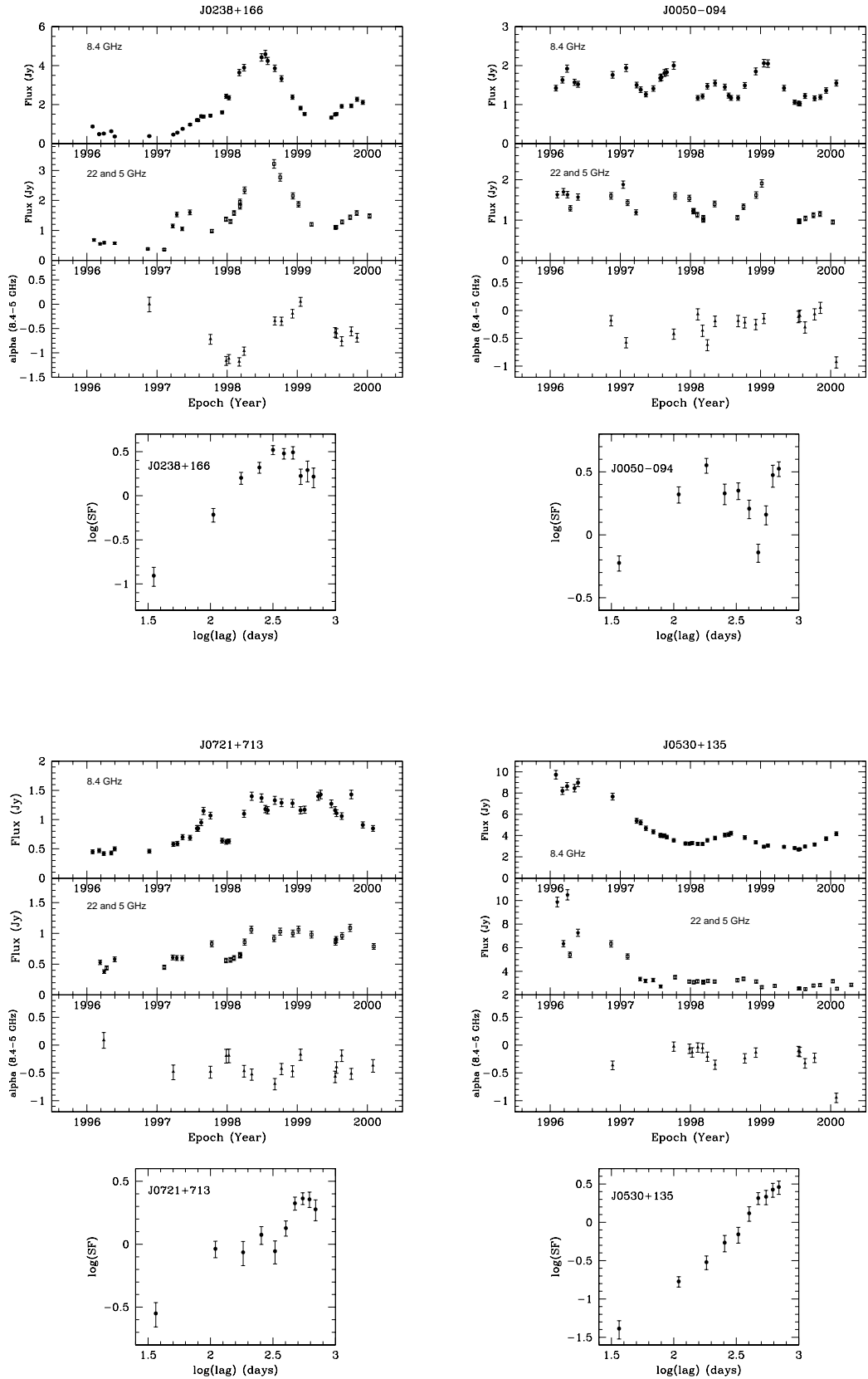}}
\caption[]{For each source: Upper panel - 8.4 GHz flux density curve data 
(upper frame); 22 GHz and 5 GHz flux density curves, shown respectively as 
filled pentagons and open squares (middle frame); 
spectral index $\alpha_{5}^{8.4}$ 
(lower frame).
Lower panel - Structure function derived on the basis of the 8.4 GHz
data}
\label{j0050}
\end{figure*}
\setcounter{figure}{0}
\begin{figure*}
\resizebox{\hsize}{!}{\includegraphics{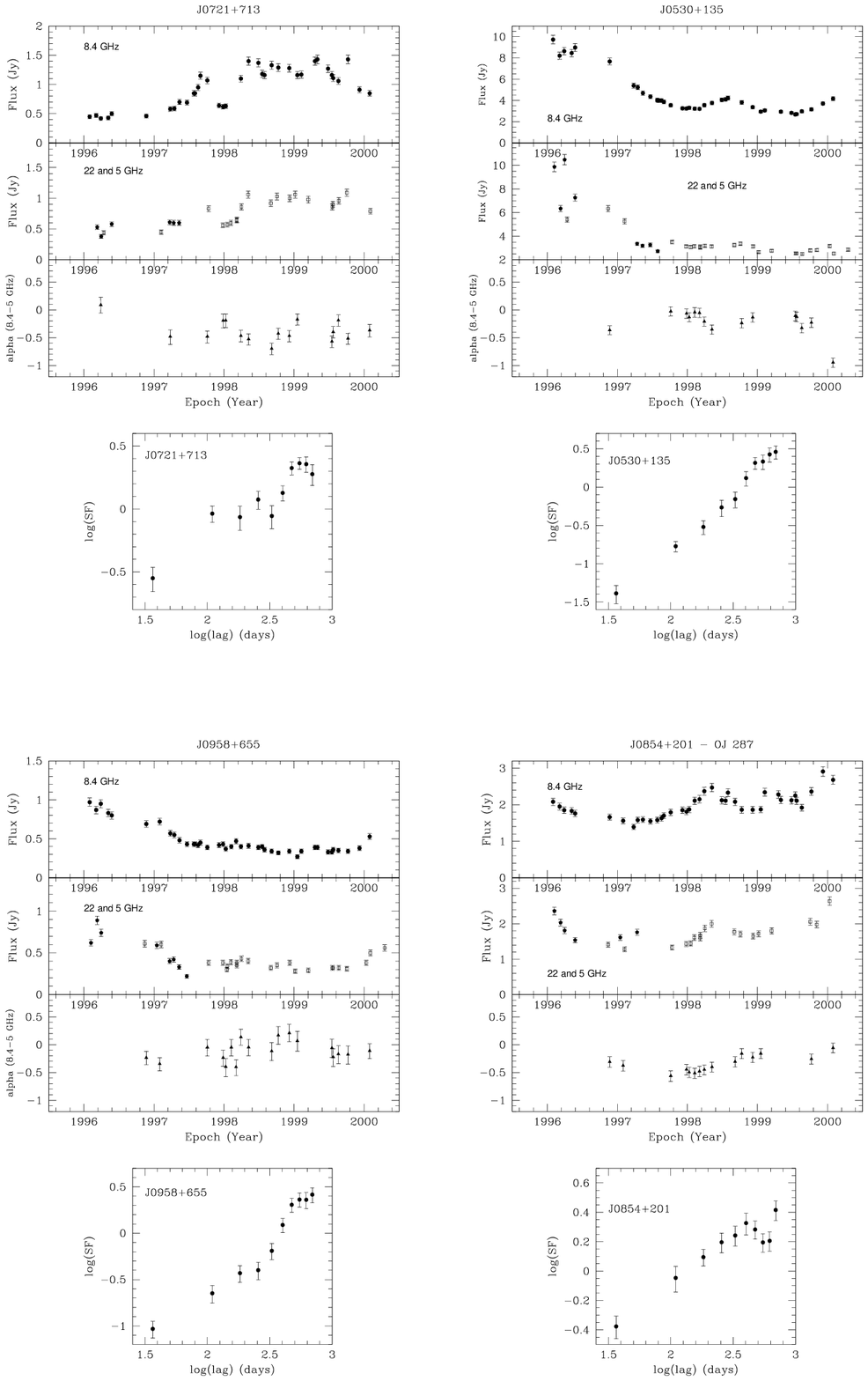}}
\caption[]{Continued.}
\label{j0238}
\end{figure*}
%
% figure 1 
\setcounter{figure}{0}
\begin{figure*}
\resizebox{\hsize}{!}{\includegraphics{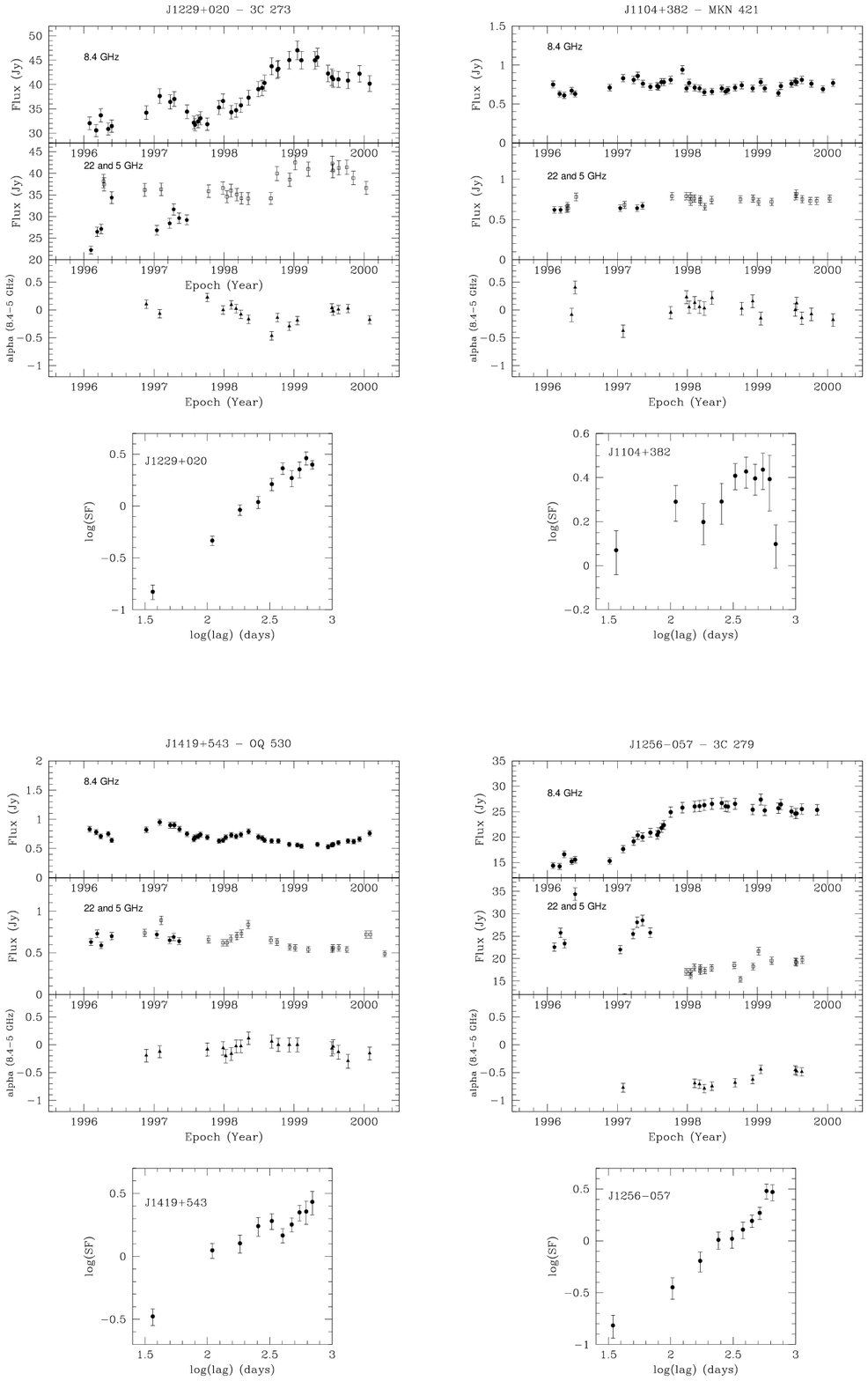}}
\caption[]{Continued.}
\label{j0530}
\end{figure*}
\setcounter{figure}{0}
\begin{figure*}
\resizebox{\hsize}{!}{\includegraphics{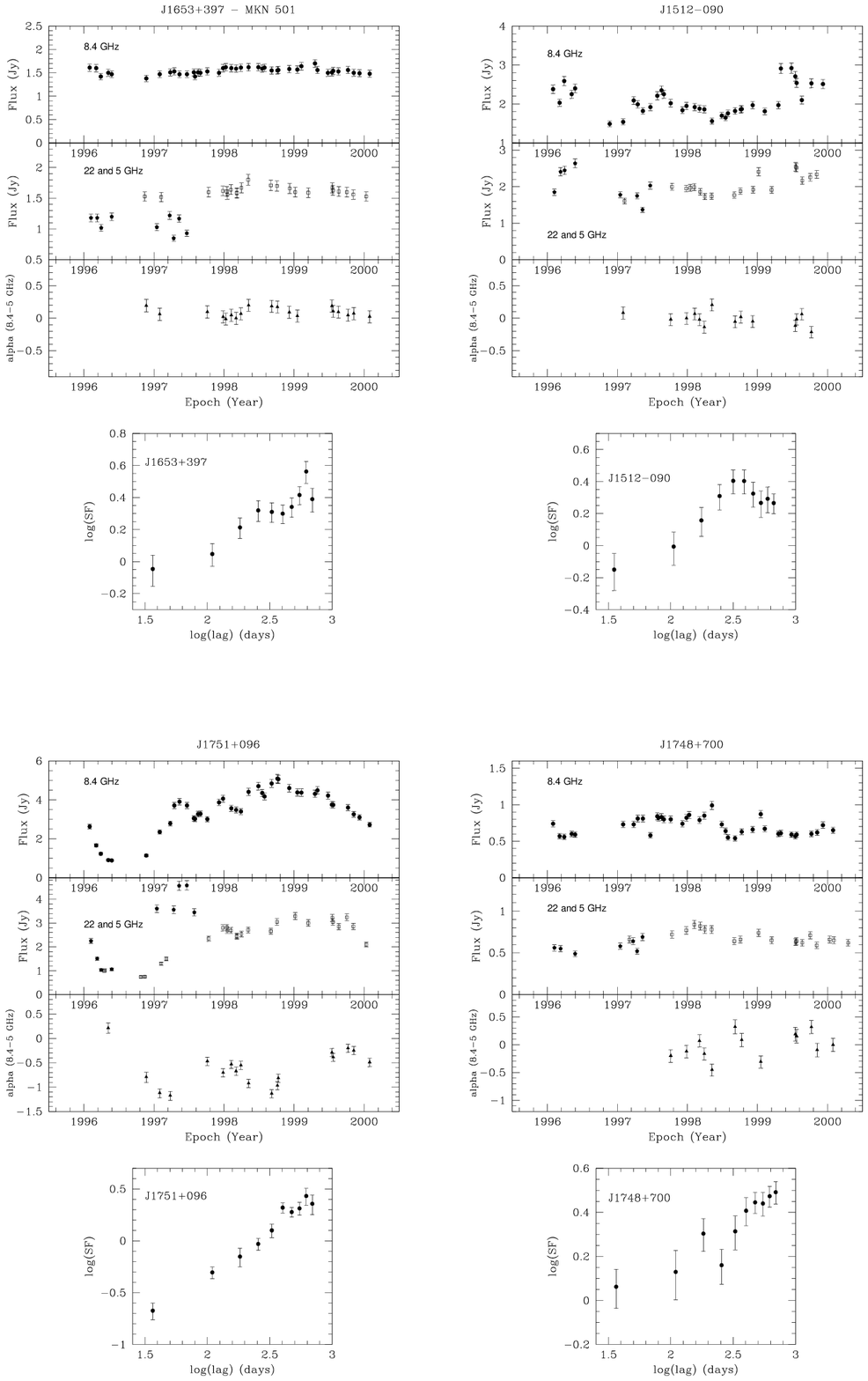}}
\caption[]{Continued.}
\label{j0721}
\end{figure*}
%
% figure 1 
\setcounter{figure}{0}
\begin{figure*}
\resizebox{\hsize}{!}{\includegraphics{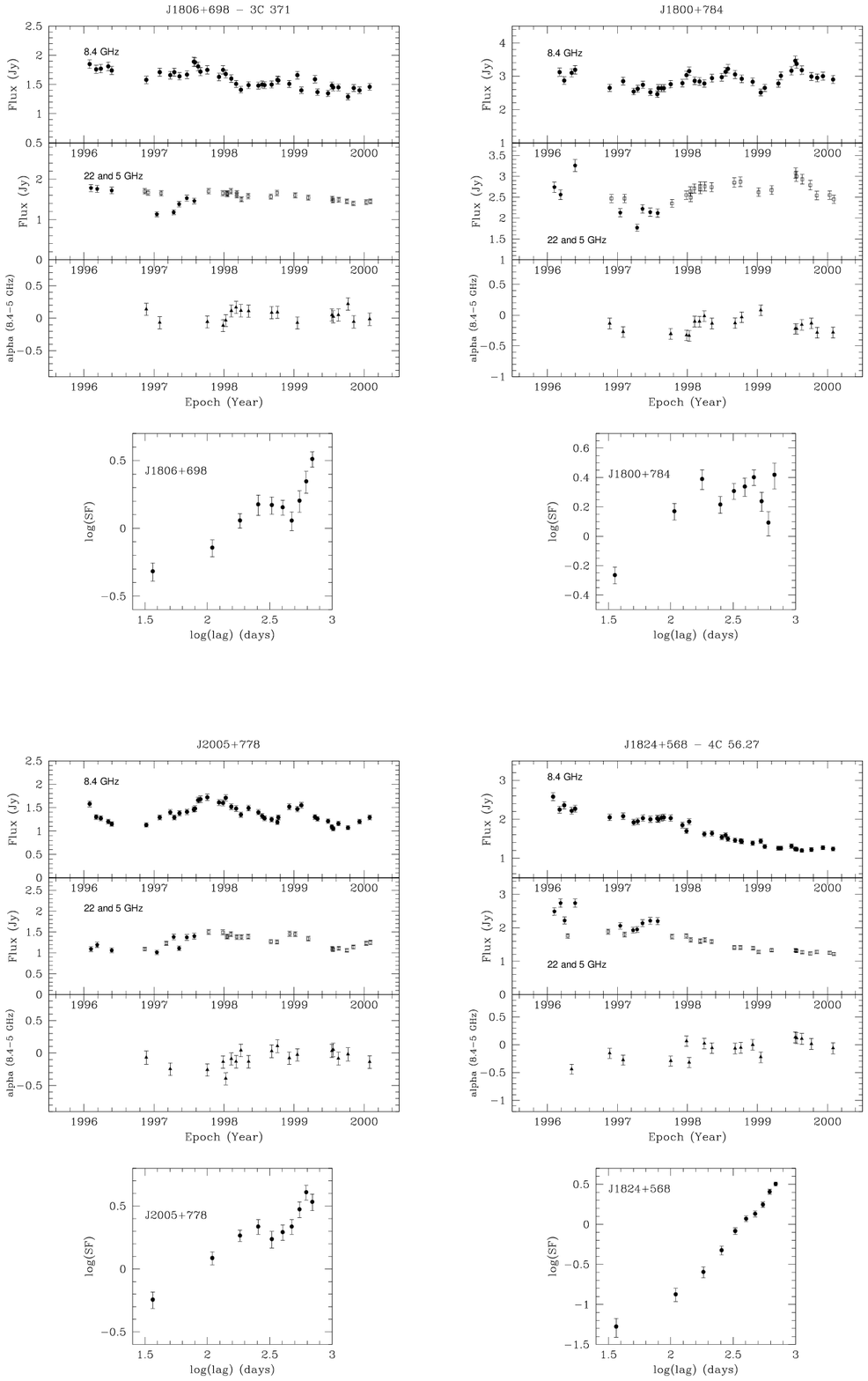}}
\caption[]{Continued.}
\label{j0738}
\end{figure*}
\setcounter{figure}{0}
\begin{figure*}
\resizebox{\hsize}{!}{\includegraphics{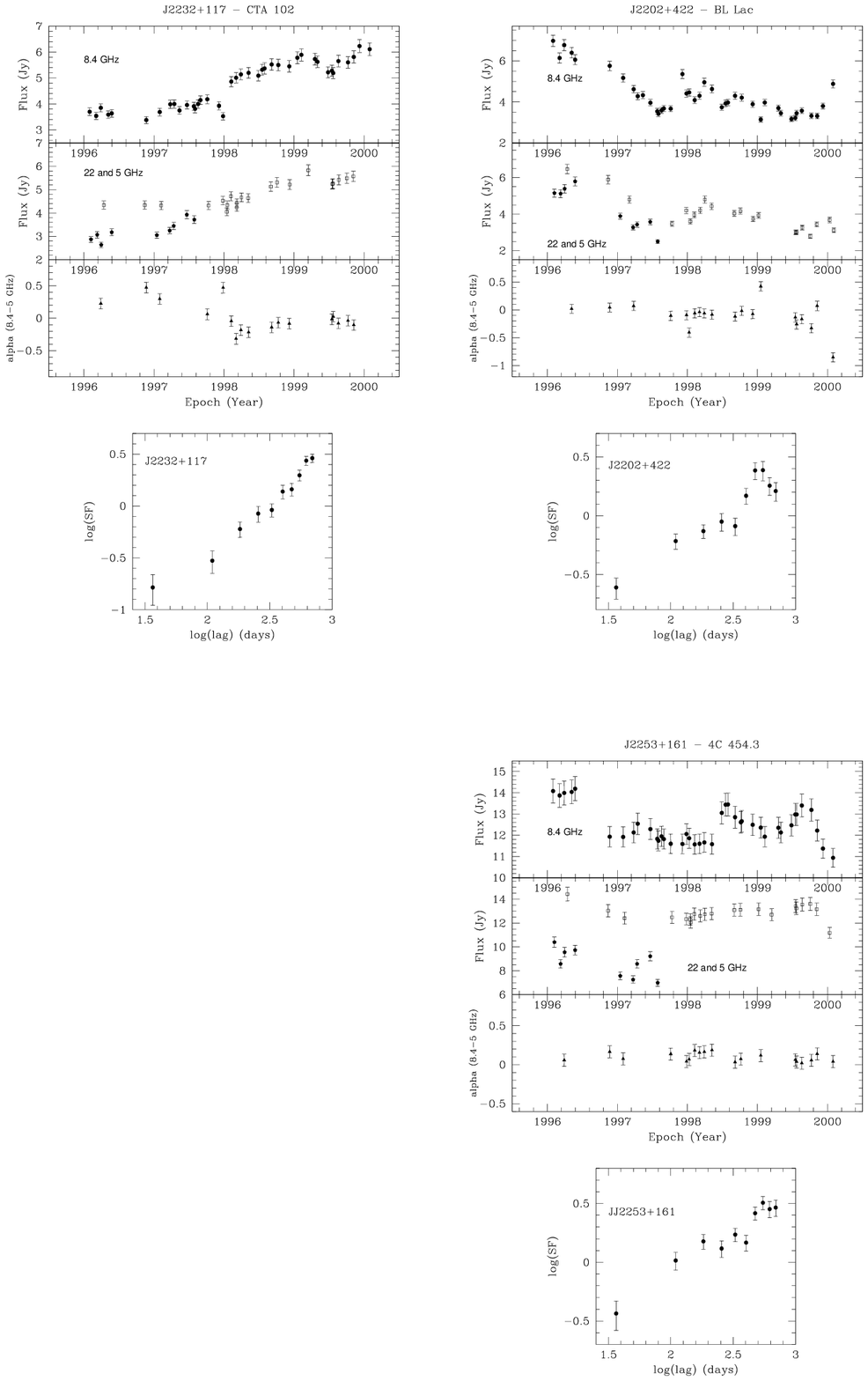}}
\caption[]{Continued.}
\label{j0841}
\end{figure*}

%%%%%%% fin qui

\end{document}